# Design, Application and Evaluation of a Multi Agent System in the Logistics Domain

J. Fischer, C. Lieberoth-Leden, J. Fottner and B. Vogel-Heuser

*Abstract—* The increasing demand for flexibility in automated production systems also affects the automated material flow systems (aMFSs) within these systems and, thus, demands reconfigurable systems. However, the centralized control concept usually applied in aMFSs hinders easy adaptation, as the entire control software has to be re-tested when sub-parts of the control are manually changed. As adaption and subsequent testing are time-consuming tasks, concepts are required for splitting the control from one centralized node to multiple, decentralized control nodes. Therefore, this paper presents a holistic, agent-based control concept for aMFSs, whereby the system is divided into so-called automated material flow modules (aMFMs), each controlled by a dedicated module agent. The concept allows reconfiguring an aMFS consisting of heterogeneous, stationary aMFMs, during runtime. Furthermore, it includes aspects such as uniform agent knowledge bases through metamodel-based development, a communication ontology considering different information types and properties, strategic route optimization in decentralized control architectures and a visualization concept to make decisions of the module agents comprehensible to operators and maintenance staff. We performed the concept evaluation using material flow simulations and a prototypical implementation on a lab-sized demonstrator.

*Note to Practitioners—* Currently, the adaption of automated material flow systems (aMFSs) concerning their layout requires modifications to the control software, including extensive testing. This conflicts the demand for flexible and reconfigurable aMFSs due to changing requirements throughout a system's life cycle. A promising approach is the modularization of aMFSs, including their control to ease layout changes and adaptations to changing material flows (known as Plug & Produce in the scope of Industrie 4.0). However, common concerns when implementing agent-based control are the real-time capability of the controlled systems and the trust of operators in the automation regarding safety and security of these systems. More precisely, it is feared that operators might be unable to distinguish the bene- and malevolent behavior of an aMFS if agents make decisions autonomously. The concept we present here addresses these issues by establishing different communication types classified according to varying real-time requirements, and by supporting operators via a human-machine interface to make agent decisions comprehensible.

## I. INTRODUCTION

AUTOMATED material flow systems (aMFSs) resemble a subgroup of automated production systems (aPSs), which transport goods from a source to a predefined sink. They are utilized in production supply, warehousing and commissioning as logistics systems within production sites and are in operation for several decades [1]. Usually, aMFSs handling piece goods are specialized for a specific type of transport unit (TU), such as pallets or small load carriers. A recent study confirms a high reuse rate of parts of an aMFS (e.g., roller conveyors) from a functional point of view: five types of conveying modules cover 95.4 % of aMFSs transporting small load carriers [2]. Still, there are as yet no standardized components or modules for aMFSs. Instead, there is a great variety of heterogeneous modules on the market, which can be combined to form a desired aMFS layout offering the required capabilities.

Furthermore, present-day aMFSs are mostly operated by a single, central control node such as a Programmable Logic Controller (PLC). Developing the control software demands manual effort, and, thus, its adaptation due to changing requirements entails extensive testing of the modified software [3]. However, demands for flexibility in aMFSs increase due to the high frequency of new or changing products, which require different operations or modifications to the aMFS layout.

Modularization approaches are a promising technique to deal with these changing requirements, where the aMFS hardware and corresponding monolithic control software are split into independent automated material flow modules (aMFMs). An aMFM is defined as an encapsulated unit consisting of hardware and the respective control software. It performs predefined logistical functions, such as transporting, buffering or identifying a TU, and can communicate with other aMFMs via standardized software interfaces and property descriptions [4]. The advantages of modularization are reduced software complexity and eased reconfigurability [5]. This approach enables modifications to the aMFS layout and the material flow abilities, respectively, by adding, removing or exchanging the aMFMs constituting the aMFS. Within the scope of Industrie 4.0, Multi Agent Systems (MASs) are considered as a suitable

This paragraph of the first footnote will contain the date on which you submitted your paper for review.

This work was supported by the German Research Foundation (DFG) with the project numbers (GU 427/25-1, VO 937/24-1).

The authors J. Fischer and B. Vogel-Heuser are with the Technical University of Munich, Chair of Automation and Information Systems, Boltzmannstr. 15, 85748 Garching, Germany (e-mail: {juliane.fischer; vogel-heuser}@tum.de).

The authors C. Lieberoth-Leden and J. Fottner are with the Technical University of Munich, Chair of Materials Handling, Material Flow, Logistics, Boltzmannstr. 15, 85748 Garching, Germany (e-mail: {lieberoth-leden; j.fottner}@tum.de).

approach for implementing such a decentralized, flexible and modular control architecture as introduced above. Thereby, software agents control parts of a system, such as the aMFMs, which communicate and collaborate to fulfill a given automation task jointly [6].

To minimize system downtime, adaptations to an aMFS should be accomplished during runtime. This is achievable if system changes are detected and configured automatically. Thus, the control of an aMFS should be able to incorporate changes in the system during runtime and to adapt to new layouts and material flows automatically. Previous works demonstrated that autonomous self-controlled aMFMs allow for reconfigurable aMFSs changeable during runtime [7].

The scope of this paper is the control of stationary aMFMs such as conveyors, while automated guided vehicles (AGVs) are not explicitly targeted. This results from differences between stationary and mobile aMFMs, e.g., AGV systems assign a transport to an AGV, which then determines a route or a path to the destination. Whereas, aMFSs with stationary aMFMs perform a transport by transferring a TU from one aMFM to another. Consequently, in the case of aMFSs with stationary aMFMs, the control determines and coordinates the sequence of aMFMs a TU passes, from a start to a destination aMFM. The control concepts differ from each other in terms of information to be exchanged, tasks to be coordinated and requirements on communication, e.g., continuously exchanging position information in AGV systems, which is not required for stationary aMFMs.

In the following, we introduce an agent-based control concept for reconfigurable aMFSs. As a first step, Section II introduces the state of the art and, subsequently, Section III sums up the problem we aim to address. Section IV presents the derived requirements for the agent-based control of aMFSs. We introduce the developed concept in Section V, followed by an evaluation in Section VI. Section VII presents a discussion and outlook of the insights gained and, finally, the paper concludes with a summary in Section VIII.

## II. STATE OF THE ART

The following section introduces the domain of aMFSs, including MASs. Furthermore, we present current approaches to increase the flexibility and reconfigurability of aMFSs and optimization strategies regarding the routing process in aMFSs.

### A. Introduction to the Domain of aMFSs and MASs

Generally, aMFSs from the domain of logistics are specialized for the transport of goods. They are mechatronic systems usually controlled with PLCs, programmed in accordance with the five languages defined in the standard IEC 61131-3, and following a centralized control architecture. Thus, modifications to the control software usually require extensive and time-consuming tests of the entire software, including unchanged parts. Model-based approaches for software development have been promoted as a solution to deal with the task complexity in the domain of aPSs [8]. To ease the development of control software for aMFSs, the conventional, hierarchical control model "System Architecture for Intralogistics" (SAIL), which is inspired by object-oriented programming, has been introduced [9]. It proposes the function-oriented modularization of aMFSs and defines standardized interfaces and logistical functions in order to describe aMFMs. However, due to its hierarchical character, SAIL is neither intended nor suitable for a decentralized control approach of aMFSs enabling reconfiguration during runtime. Yet, due to their modular hardware setup, aMFSs are suitable for implementing an agent-based control concept. They consist of aMFMs, which resemble autonomous entities, each able to execute predefined tasks. These aMFMs can be realized utilizing agents that control one aMFM each, and communicate with other aMFM agents to accomplish tasks within an MAS.

A renowned specification regarding the overall design of MASs is the agent management model defined by FIPA (Foundation of Intelligent Physical Agents), which develops suggestions and standards for MASs [10]. It provides the framework for agents to exist and operate in and, furthermore, includes communication concepts for agents. The main components of the framework are the agent platform (AP) with an agent management system (AMS) to register the agents, a directory facilitator to enable the search for agents and their capabilities within the MAS and, finally, the agents themselves. A dedicated message transport system enables the communication between agents within and across an AP. While the framework introduces the main concepts and their abilities and responsibilities, the implementation details of individual APs and their agents are not specified. Recently, attempts in the automation domain to standardize the utilization and design of agent systems by providing guidelines and best-practice rules as well as application examples have been enforced [11].

Within the scope of MASs, different communication strategies between agents have been analyzed, e.g., Ulewicz et al. distinguish between two basic types of information exchange to deal with the partly conflicting requirements of real-time capability and flexibility [12]. The first method of information exchange is direct communication between agents via messages, which is suitable for complex, highly dynamic interaction. Secondly, they propose local coordination between agents for time-critical, cyclic information exchange, to enable real-time communication required by aPSs.

Whenever software agents make decisions without the intervention of human operators, the choice is like a black box and not necessarily comprehensive for the operator. This can jeopardize the operator's trust in the correctness of the decision of the MAS. Lee et al. defined trust as "the attitude that an agent will help achieve an individual's goals in a situation characterized by uncertainty and vulnerability" [13]. The user's trust in automation is directly linked to the degree of reliance on a complex automated system. Both, over- and under reliance on such a system can have a negative impact on the proper functioning of the system [14]. Thus, when designing MAS, this aspect needs to be considered.

### B. Approaches to Enable Flexible aMFSs

To increase the flexibility of manufacturing machines, and to enable modifications while avoiding error-prone, manual adaptation processes dependent on the operator's knowledge, Marks et al. [15] present an agent-based decision support system. The MAS supports the user by generating and evaluating adaptation options to meet requirements for new production requests during a machine's life cycle, and an estimation of the modification effort. Its decisions are based on a product, process and resource model of the considered machine, which contains parameter ranges, interdependencies between system parameters and a given production request. Thus, the focus is on the generation of change and adaptation possibilities, with the MAS functioning as a decision system for planning purposes, but not for the actual control of the system.

Similarly, Beyer et al. present an MAS approach for designing and choosing a suitable layout for an aMFS [16]. The development engineer enters a set of requirements for the aMFS to be designed in an assistance system, which generates alternative design solutions and rates them according to the specified requirements. This approach can be applied when developing new aMFSs, and for reconfiguring or optimizing existing aMFSs, but does not consider the control of aMFSs. Libert presents a procedure for the development of an individual, agent-based material flow control for aMFSs [17]. As an agent platform, he uses the PC-based JADE-Framework and does not apply his concept to industrial real-time control hardware, such as a PLC.

Several approaches have introduced standardized modules to build reconfigurable aMFSs controlled by MASs, with either a centralized or decentralized control. Black et al. developed an MAS for baggage-handling systems using IEC 61499 Function Blocks, where each block represents a module [18]. Routing conflicts are avoided thanks to the system layout. Approaches that rely on completely centralized control hardware, such as [19], lack flexibility in terms of scalability and lack redundancy for convertible aMFSs. Priego et al. present an agent-based approach for the reconfiguration of an IEC 61131-3 software program assuring availability during runtime, in the case of a controller failure [20]. For this purpose, the system under control is divided into mechatronic components (MCs), and the software to control a specific MC is replicated on different controllers. Each MC contains, additionally to the actual control software, the information required for reconfiguration, i.e., the relocation of an MC control software to another controller. Thereby, a metamodel defines the reconfiguration requirements. However, the approach focuses on the control hardware and does not consider system reconfigurations due to the addition or the removal of a module to or from the system. Kovalenko et al. present a methodology to enable flexible manufacturing systems based on two different kinds of agents [21]. Additionally to the commonly deployed resource (aMFM) agent, an intelligent product (TU) agent is implemented that allows the product to explore the capabilities of the surrounding manufacturing environment. However, the approach focuses on the communication and collaboration of different agent types and does not consider reconfiguration of the controlled system.

Vallée et al. introduced an automation agent to describe a module in a modularized production system [22]. The automation agent is divided into a physical and a software component, which is further divided into an upper and lower control level. Thereby, the lower control level facilitates a limited variety of abilities and reads/writes hardware in- and outputs in real-time. The upper control level, implemented in accordance with FIPA [10] or similar, represents the module in the MAS, coordinates tasks with other agents and pursues the module's targets. The design pattern is applied in several concepts for flexible production systems [23] but has not yet been applied to aMFSs.

To increase the aMFSs' flexibility, Mayer developed the hardware for autonomous material flow conveyor modules and the software for a decentralized routing and reservation process to avoid deadlocks [24]. The approach is limited to a single type of conveying aMFMs without considering manipulation aMFMs. Furthermore, no central coordination instance is provided. Modules are connected through predefined hardware ports, and thus detect the position of their neighbors.

Transport costs are a primary influence on the routing decision. Thus, traffic management can be applied by modifying the costs of transport on a module [25]. However, especially in complex systems, it is challenging to optimize traffic through link (aMFM) costs, because changing the costs of one module might cause unwanted traffic situations in other system areas. In communication networks, the Multi-Label Protocol Switching (MLPS) provides a flexible solution for traffic engineering [26]. Routes are calculated online or offline and capacities are optimized. Afterward, the affected network routers are informed. A route is selected and assigned to a data package and routers forward the data package along the designated route. MLPS was developed for the virtual transport of data packages and has not yet been applied to aMFSs.

The main contribution introduced in this paper is an agent-based field-level control approach (IEC 61131-3-compliant), enabling autonomous reconfiguration for aMFSs during runtime while taking into account optimization strategies and the operators' trust in the MAS's decisions (cf. [27] for challenges to overcome regarding autonomous systems).

## III. PROBLEM STATEMENT

In the scope of Industrie 4.0 and highly customized products with lot size one, there is an increasing demand for flexible, reconfigurable aMFSs, which adapt to changing requirements during runtime. However, to reduce the downtime of aMFSs, layout changes are usually performed under high time-pressure, including required software adaptations, which entails a high risk of causing technical debt [1]. Although current research promotes using MASs to enable Plug & Produce, where aMFMs are added to an existing aMFS and are usable for production shortly after their addition (cf. Section II), open challenges when attempting to change the layout of an aMFS during runtime remain. While Plug & Produce requires communication between heterogeneous aMFMs and a suitable strategy for route optimization, the use of an MAS demands measures to establish the operator's trust in an MAS's decisions. From current challenges and state of the art, we derived requirements for an agent-based control concept for aMFSs, which are presented in the subsequent Section IV.

## IV. Derived Requirements for MASs in aMFSs

The following Section presents requirements for an agent-based control of aMFSs, which are summarized in Table I.

### A. Metamodel-based Description of aMFMs

In the scope of Industrie 4.0, the concept Plug & Produce – where production modules can be added and used for production shortly after – is a current subject of discussion. To enable the flexible control and reconfiguration of aMFSs similar to this concept, a uniform description of aMFMs is required. Generally, the addition of an aMFM to an aMFS comes with the registration of the respective module agent to the MAS regardless of its type (i.e., a roller conveyor or a portal crane) or the abilities (e.g., transportation or manipulation) it offers. Therefore, the aMFM description, which forms the knowledge base of the module agents, needs to be modeled uniformly. Furthermore, aMFM agents need to interact and communicate with each other to jointly perform the required transportation tasks, regardless of their type. For this purpose, uniform interfaces are required. Uniform interfaces are also used to derive connections between neighboring aMFMs and, therefrom, routes through the aMFS based on its current layout.

Additionally, a module agent's knowledge base needs to contain information on the aMFM's capabilities to determine if the respective aMFM can perform a requested transportation or handling order. Finally, to execute a given transportation order, a module agent needs detailed information on the aMFM hardware it controls. To fulfill the need for a consistent and uniform description of different aMFMs, a metamodel-based development of the agent's knowledge base can be applied. Thus, the first requirement regarding the agent-based control of aMFS is as follows: To enable the communication and collaboration between agents independent of their aMFM type, the knowledge base of aMFM agents needs to be uniform, and thus defined based on a metamodel (R1).

### B. Communication Ontology for Data Exchange Between aMFM Agents

Within aMFSs, various tasks need to be performed, which require interaction and information exchange between the aMFMs within the system. Depending on the given task, the amount of information to be exchanged, the number of entities communicating and the requirements regarding the actuality of the information all vary. For example, to derive the current topology of an aMFS, dimensions, including interfaces and localization data of all aMFMs within the system, are required. As stationary aMFMs are considered exclusively, only localization information of newly added or (re)moved aMFMs needs to be updated, while the remaining aMFMs and their localization data are kept. Thus, only the event of layout modifications requires an update of the localization data of added or removed aMFMs, which the AMS registers centrally. Concerning planning, scheduling and deciding on a route through the aMFS for a given transportation order from a superordinate system (e.g., warehouse management system), the current topology of the aMFS including connected aMFMs is required. Subsequently, the fine-grained routing task involves multiple aMFM agents negotiating which aMFM can perform which (sub-)task. Usually, this route planning is conducted in advance and does not require real-time communication. But in the event of material transfers from one aMFM to its connected neighboring aMFM during the execution of a scheduled route, communication between the two respective aMFM agents is time-critical, if the handover requires a synchronized control of the aMFM hardware, as it is often the case. These examples demonstrate that the amount and type of data to be exchanged and the communication properties vary greatly.

Overall, the communication needs to be enabled regardless of the aMFM type and it should be scalable regarding the amount of communicating agents in the aMFS. Furthermore, a general communication ontology is not feasible as the data to be exchanged varies greatly. Therefore, a communication ontology needs to be derived, taking the different interaction scenarios within an aMFS into account. This includes a classification of the exchanged information and communication methods to be applied, as proposed by Ulewicz et al. [12]. Thus, requirement 2 demands different communication ontologies for different types of information (R2).

### C. Routing Process and Its Optimization

Regarding the routing process within aMFSs, optimization and preventing deadlocks and blockages play an essential role in enabling high throughputs. Optimized routes increase throughput and decrease the average process time of transports. While optimization strategies for aMFSs being controlled by a central control node such as a PLC have been subject to study for decades, the application of decentralized optimization strategies is not yet common. To enable flexible, reconfigurable aMFSs, however, this paper targets an agent-based control architecture and distributes the control across decentralized aMFM agents. Thus, existing strategies need to be adapted to be applicable. Decentralized or distributed controlled, highly flexible routing concepts for aMFSs determine an optimal route for one specific transport for the current state shown in Fig. 1, left. Highly flexible routing causes a high communication load with regard to traffic. For each transport task, one aMFM must collect the current system information, or the routing task is distributed over numerous aMFMs, even though the final route does not affect them. However, strategic route optimization aims to find a route for a transport, which leads to an overall long-term optimization of all transports. Thereby, individual transports might not be assigned an optimal route and take longer, but the overall performance in terms of throughput and average process time improves. In contrast to highly flexible routing, static routes can be implemented in an aMFS. Static routes are determined (once) either manually by the developer of the aMFS, or automatically calculated. Each transport uses the same static route, so no communication and calculation for routing is required during the operation of an aMFS. However, static routes do not react to layout or material flow changes and are, thus, inefficient to control flexible aMFSs. An adaption of static routes are semi-static routes (ssR) similar to the MLPS concept used in communication networks [28] shown in Fig. 1, right. In communication networks, MLPS has been introduced to enable traffic management for strategic optimization. The concept of MLPS can also be applied to routing in aMFSs.

TABLE I
OVERVIEW ON DERIVED REQUIREMENTS

| No. | Derived Requirement |
|---|---|
| R1 | To enable communication and collaboration between agents, independently of their module type, the agents' knowledge base needs to be metamodel-based |
| R2 | The communication ontology needs to support different types of information with varying demands on real-time. |
| R3 | A concept to enable strategic optimization based on semi-static routing is required for agent-based aMFS control. |
| R4.1 | Every aMFM needs to be equipped with its own visualization to enable a visualization of the current aMFS layout, including the present aMFMs and their connections |
| R4.2 | The visualization of the current aMFS layout needs to be enhanced by relevant parameters to enable the operator to understand the decisions of the MAS. |

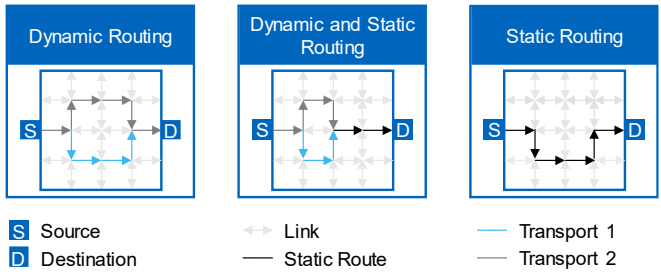

Fig. 1. Comparison of dynamic and static routing concepts.

Highly flexible and ssRs can also be combined into one system. The concept of ssRs can be applied in critical areas, where multiple transports take place and optimization is required, i.e., has a great influence on the performance, which is shown in Fig. 1 middle. The concept of highly flexible routing is applied in areas with irregular, many different and only few transports. The combination of these two routing concepts decreases the demand of ssRs. On the other hand, the complexity of the control increases because two different routing concepts are incorporated into the same aMFS and an additional concept is required to determine which area is controlled by flexible and which by ssRs. Therefore, in this paper, we only pursue the concept of ssRs to limit the complexity of the control. Thus, a concept to enable strategic optimization based on semi-static routing is required (R3).

*D. Visualization of aMFS Layout and MAS Decisions*

In the life cycle of an aMFS, a so-called layout plan including aMFMs, their positions within the system and connections in between them from a bird's eye perspective is used from an early stage. Thus, the layout plan provides a reasonable basis for a human-machine interface (HMI). As the layout in a flexible, reconfigurable aMFS is subject to changes throughout the system's operation phase, the HMI correspondingly needs to adapt to display the current aMFS layout. Since the aMFMs to be added to the aMFS cannot be foreseen, a flexible concept for the HMI is required. Such a concept enables the visualization of the current aMFS layout if every aMFM is equipped with its own module visualization (R4.1).

While the operator will gain an overview of the aMFS through this HMI, information regarding the decisions of the MAS is still missing. However, a common concern when dealing with the implementation of agent-based control in automated production systems is the interaction between the operator and the control system. More precisely, the operator needs an appropriate level of trust in the MAS decisions [13] and to be capable of distinguishing normal behavior of the aMFS and behavior indicating a malfunction, i.e., the operator needs to recognize situations, which require his interference due to safety or security threats. Thus, the decisions of the MAS need to be visualized, including factors or parameters, which are significant for the decision, such as selected parameters of the material flow, the status of individual aMFMs inside the system and other parameters influencing the MAS's decisions. Overall, the aMFS layout visualization needs to be enhanced with additional information that enables operators to understand the decisions of the MAS regarding transportation routes and to distinguish between bene- and malevolent situations (R4.2).

The derived requirements are summarized in Table I.

V. DEVELOPED AGENT-BASED aMFS CONTROL CONCEPT

In the following, we present the developed concept for agent-based control of aMFSs with a focus on the fulfillment of the requirements introduced above. To enable the agent-based control of aMFSs, a function-oriented modularization approach is applied [29]. Each aMFM, performing either a transportation or a manipulation ability, is controlled by a dedicated PLC, programmed in compliance to the IEC 61131-3. The respective control software consists of two parts, namely the pure hardware control of the aMFM's automation hardware, and an agent-part. We separate the agent part into two sub-parts, which are the so-called aMFM agent and the agent framework. The aMFM agent represents the aMFM in the MAS and has only a limited view over the aMFS consisting of a priori implemented knowledge, information from the aMFM's sensors and information exchanged with other aMFMs. While the aMFM agent is acting locally and is designed aMFM-specific, the framework operates globally and is identical on every PLC. It functions as the central instance in the MAS and registers, following FIPA's AMS, all aMFMs that are added to the aMFS. It also serves as an interface to higher-level systems, such as a warehouse management system. Furthermore, the framework creates a consistent aMFS topology with little communication effort by collecting and merging the local aMFM topologies. Although the framework is present on every PLC, it is only active on one of them. Overall, the chosen architecture is a compromise in which the framework serves as a redundant, central instance to reduce the communication load and to provide consistent information concerning route determination. Generally, a central, redundantly designed instance represents a limitation in scalability if, for example, the maximum communication or computing load is reached. For this reason, the framework is limited to one-time processes (e.g., configuration process for a newly added aMFM) or to collecting and providing consistent data as far as possible and named coordinator in the following. Detailed planning and calculation processes are distributed among the aMFMs (cf. [7] for details).

*A. Metamodel-based Description of aMFMs*

Based on the SAIL-compliant metamodel *AutoMFM* [4], heterogeneous aMFMs can be modeled, and thus the respective aMFM agents' knowledge base is formed in a uniform manner. The metamodel includes several aspects, such as information on an aMFM's general description, its status, its abilities and dimensions, and the control of its actuators (cf. Fig. 2).

The metamodel also contains information on an aMFM's material flow properties, by specifying the material handover required when transferring or receiving a TU via its physical interface. However, the metamodel holds only one type of aMFMs without a further distinction into transportation and manipulation aMFMs. While transportation aMFMs are required for the actual transportation task, manipulation aMFMs offer additional abilities, e.g., labeling. Thus, the simplified metamodel *IntraMAS*, which is similar to *AutoMFM* and supports different aMFM types required for the agent knowledge base, was established [30]. *IntraMAS* defines the structural composition of aMFSs, including routes and orders.

The agents' knowledge base consists of two parts: information regarding the controlled aMFM itself, such as the geometry or abilities offered by the aMFM (which is referred to as static information) and knowledge such as current, connected neighbors of the aMFM (which represents dynamic information calculated during runtime and exceeds the initial knowledge of the aMFM agent). The software developer creates the static information, such as the aMFM description, once during the development of the aMFM's control software in IEC 61131-3. For this purpose, the developer should use the information available from other models, such as the dimensions of an aMFM from a CAD-model or the mechanical layout plan of the aMFS (cf. [30] for details). To support the exchange of geometrical or behavioral data between heterogeneous models such as CAD-models, the data exchange format AML was developed. However, despite the use of available data from existing models, the initial effort for the knowledge base development of an aMFM type is quite high. Nevertheless, the high degree of reuse of aMFM types in aMFS systems allows the assumption that the initial effort for the knowledge base creation pays off [2].

Knowledge exceeding the aMFM itself and requiring interaction and information exchange with its surroundings, such as the current aMFS layout, interfaces to other aMFMs and available routes, is calculated during runtime of the aMFM and provided to the module agent, e.g., by the coordinator.

*B. Communication Ontology for Data Exchange Between aMFM Agents*

Within an aMFM, the information to be exchanged and processed is identified and divided into two communication categories. These include information with a high demand on timeliness (e.g., information during the physical execution of an operation) and planning information, where an unexpected delay does not interfere with the aMFM task (e.g., information on the utilization of an aMFM) or even lead to an unsafe state of the aMFM. The developed communication ontology includes both communication categories. It applies the approach of Vallée et al. to distinguish between a reactive lower control level for timeliness and a proactive upper control level for planning processes in the presented concept for aMFSs [22]. Combined with the coordinator agent, it enables performing a global planning and coordination assignment, while locally deploying a conventional logic control.

The tasks of an aMFM's upper control level can be further divided into independent sub-tasks on separate logic levels, resulting in a hierarchical agent with different logic levels. A hierarchically structured agent avoids conflicts of interests because the decision of a lower logic level has to be in accordance with the decisions of higher logic levels [31]. Accordingly, the aMFS control represents a hybrid architecture with both, distributed and decentralized agents forming the MAS and a centralized coordinator in order to reduce communication loads and serving as an interface to superordinate systems. The higher logic level of an aMFM is in charge of avoiding blockades. When lower logic levels act in accordance with the decisions of high logic levels, the lower logic levels are by design unable to cause blockades themselves. According to the aforementioned hierarchy, an aMFM agent is divided into four logic levels (cf. Fig. 3): Material Flow, Functional, System and Configuration.

At the top, the material flow logic level is responsible for planning and coordinating the global material flow with other aMFMs. This logic level is responsible for selecting routes through the aMFS and for avoiding blockades between TUs passing the same aMFM. Thus, the material flow logic level determines the sequence in which a given TU passes the aMFMs. To enable cooperation between aMFMs, the material flow logic level is identical in all aMFM agents. The functional logic level plans and coordinates transports within an aMFM. As there are different types of aMFMs, which differ in functionality, size, control strategies, and other properties, the functional logic level is implemented individually for each aMFM type. It organizes and optimizes transports following the TU sequence determined by the material flow logic level. As a result, the functional logic level determines the sequence of operations and parameters to be executed to transport a TU through the aMFM. Finally, the system logic level represents the lowest control level. It controls the aMFM hardware by reading and writing its IOs, and executes the sequence of operations. The configuration logic level performs an MAS's organizational tasks. It is responsible for registering an aMFM in the aMFS, configuring its interfaces according to detected neighbors, and supervising it for changes. Overall, a common knowledge base and its storage space, which has a standardized structure for all aMFMs, facilitate the communication as it is used by all logic levels of the aMFM agent.

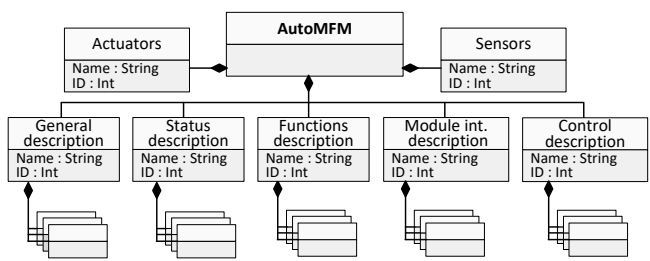

Fig. 2. Excerpt of AutoMFM to design the knowledge base of aMFM [4].

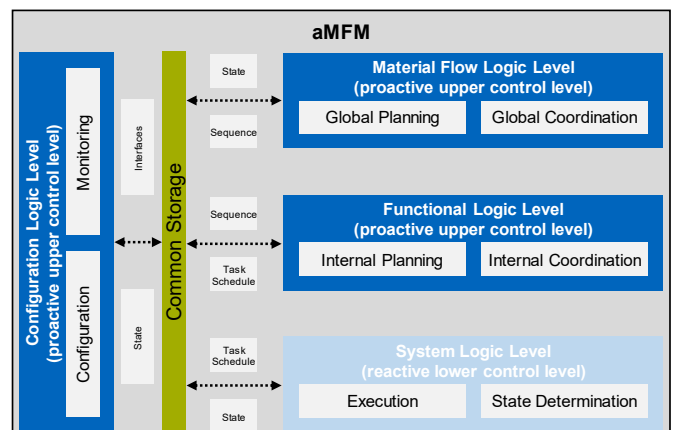

Fig. 3. Hierarchical architecture of an aMFM agent.

### C. Routing Process and Its Optimization

For each material flow relation, the ssR concept negotiates a route in the aMFS, based on the already existing routes and the required capacity, priority, etc. of the material flow relations. A material flow relation states how many TUs are transported from a system entrance (source aMFM) to a system exit (sink aMFM) and possesses properties such as average throughput. The result is a path of aMFMs through the aMFS with properties, such as a granted (reserved) capacity, and called ssR. Each aMFM, which is part of an ssR, is informed about the predecessor and successor aMFM and the properties such as the reserved capacity for the route. All TUs of the material flow relation use the same ssR. Subsequently, the aMFS is meshed up with various ssRs for each material flow relation.

For the calculation of ssRs the two target dimensions to be optimized are capacity and process time of an aMFS. In order to consider two target dimensions, constraint-based routing is applied. First, only those aMFMs are considered, which provide sufficient available capacity for the already assigned and new ssRs. Next, an aMFM path with the shortest process time from the source to the sink aMFM is determined, e.g., using the Dijkstra algorithm. As a result, a set of ssRs distributes the transports in the aMFS and avoids congestions, as the maximum capacity of each aMFM is not exceeded. If the path with the minimal process time of various material flow relations will lead through the same area and cause congestions, the constraint-based routing approach selects a path for some material flow relations, which will detour the critical area if applicable. Subsequently, ssRs adopt from MLPS the favorable feature of distributing the transports across the whole system, which leads to better utilization of the system's resources and better overall performance.

ssRs are not irrevocably set for the lifetime of an aMFS but are renegotiated if the layout or material flow relations substantially change, i.e., a material flow relation requires a higher capacity than initially reserved for the assigned ssR. In that case, an established ssR is revoked and a new route with another aMFM path or capacity is reserved for a material flow relation. ssRs indicate future transports. A set of ssRs is valid for the future as long as the material flow relations are stable (constant average throughput and variability). Stable material flow relations might occur for several months, days, hours or only minutes and depend on the task of an aMFS. The longer the material flow relations are stable, the higher the influence of optimization. Every change of ssRs causes disturbance in the aMFS and may lead to temporarily inefficient transports if some transports finish on outdated ssRs while others are already using new routes. Changing the layout in aMFS requires a manual effort in the installation or removal of aMFMs from the operator. Thus, the layout usually changes at most on a daily basis, and otherwise less frequently. In contrast, the material flow relations might alternate over the day, e.g., when an aMFS is used in the receiving and shipping area, where goods arrive in the morning and leave the site in the evening. In the interim, stable material flow relations often develop for some time. Thus, the ssR concept is feasible for aMFSs.

However, some differences in the behavior of transports in aMFSs and communication networks need to be considered:

1. TUs possess a mutual physical dependence, i.e., cannot be transported in opposite directions on the same link (aMFM) at the same time.

2. In the case of a conflict, TUs cannot be deleted and restarted from the source.

3. The capabilities as throughput and process time of an aMFM can depend on the transport properties (e.g., direction) and frequency (e.g., batch processing).

4. TUs are usually transported in a continuous flow and cannot be temporarily stored and rearranged at an aMFM if the aMFM is not physically capable to do so.

The first constraint, in particular, requires an adaption of the MLPS concept. Fig. 4 shows the effect of the first constraint. TUs are transported from left to right and right to left and can either take the upper or lower branch. Communication networks choose a branch dependent on the current occupancy of the branches. Opposing transmissions on one branch do not affect each other because of the negligible transfer time and only virtual presence of the TU. Consequently, the direction of transports does not significantly affect the performance of communication networks (isotropic behavior). In contrast, the direction of transports affects the performance of aMFSs (anisotropic behavior), especially in the example shown. If both branches are occupied with an opposing transport, a TU has to wait until a branch is cleared. Subsequently, ssRs utilize an adaption of MLPS, which incorporates the anisotropic behavior of transports in aMFSs in the calculation and optimization.

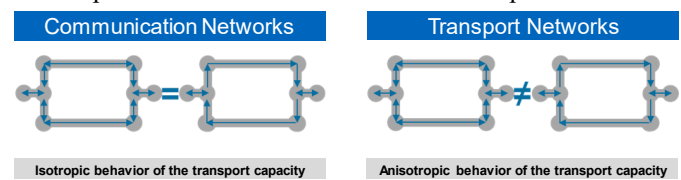

Fig. 4. The direction of transport routes does not affect the capacity in communication networks, in contrast to transport networks.

The main advantage of ssRs is the transparency of routes, on which aMFMs will transport future arriving TUs through an aMFS: The operator and the MAS can analyze the future material flow, apply various optimization concepts, and estimate and evaluate material flow after adaptions. Similar to MLPS concepts, on- and offline optimization is feasible. Furthermore, ssRs enable the visualization of routing decisions and capacity constraints for operators in aMFSs with distributed and decentralized control. It is even possible for the operator to change ssRs manually.

A predefined route for a material flow relation does not guarantee conflict-free routing, i.e., an aMFM can be reserved for two opposing routes. Furthermore, the ability to establish a sequence depends on the route and timing of a transport. Therefore, before a transport task starts, the material flow control firmly schedules TUs on every aMFM on the route, which the ssR predetermines. Firm scheduling ensures that the TUs arrive in the requested sequence at the sink. Advanced planning also avoids deadlocks, i.e., transporting two TUs in opposing directions on a bidirectional conveyor at once [32].

### D. Visualization of aMFS Layout and MAS Decisions

To provide an HMI to operators or maintenance employees, each aMFM is equipped with a simplified 2D visualization, including aspects such as the aMFM dimensions, interfaces and controllable actuators. When adding an aMFM to the system, its visualization file is manually stored on a central computer and, subsequently, automatically linked with the localization data of the aMFM, i.e., its position and orientation within the production site. Based on this, the current aMFS layout visualization (consisting of partial aMFM visualizations) is enlarged by the newly added aMFM.

To permit a focus on different aspects, the visualization includes various views: the layout can optionally be annotated with aMFM information to make clear, how many aMFMs currently compile the aMFS and which connections have been derived between them (cf. Fig. 5). Additionally, the active coordinator module can be highlighted. Furthermore, actuators that are currently running are visualized, including their direction of movement or their current orientation, e.g., the conveying direction of transportation aMFM or the orientation of the switches (cf. Fig. 5, indication of currently running conveyors and their conveying direction with arrows).

Moreover, to support the operator in understanding the MAS decisions, the current system status, including aspects such as the workload of the aMFMs can be visualized (cf. pie charts in Fig. 5). Another view highlights all available routes through the aMFS, including their properties, e.g., reserved capacity, used capacity or average duration of transportation. Thus, in combination with the currently chosen strategy for routing optimization (balanced workload, minimal cost or minimal time in the system), the operator can determine if the agents' decisions on the route are comprehensive, or if a malevolent behavior of the plant is suspected.

An additional view in the HMI provides a list of scheduled orders and the chosen routes to perform the individual orders.

## VI. Evaluation of Agent-based aMFS Control

For evaluation purposes, different means are used, ranging from material flow simulations to an application on a lab-sized demonstrator. The following subsections present the results in correspondence to the requirements.

### A. Metamodel-based Knowledge Base of aMFM Agents

To evaluate the metamodel used for a uniform definition of the agents' knowledge bases, in a first use case, we use a PLC-based simulation combining a transportation and a manipulation aMFM in the form of a portal crane. In the evaluation scenario "increase throughput" of the considered aMFS excerpt, a second portal aMFM is successfully added during runtime of the system and, subsequently, incoming transportation orders are distributed between the two portals without manual interactions [30]. Thereby, the knowledge base of the module agents and their uniform aMFM descriptions are sufficient to control the scenario. In a second evaluation, the same metamodel is used as a basis to form the knowledge base of agents controlling aMFMs within the Industrie 4.0 demonstrator MyJoghurt [33]. For this purpose, we divided the demonstrator's logistics part into four aMFMs, each controlled by a Beckhoff PLC (cf. Fig. 5). Overall, we evaluated four scenarios, namely:

1. Start-up and system initialization
2. Addition of an aMFM during runtime
3. Removal of an aMFM during runtime
4. Removal of active coordinator aMFM during runtime

To add an additional aMFM to an aMFS, first, an operator links the aMFM hardware at the desired location. Subsequently, the respective PLC controlling the aMFM is connected to the aMFS PLC network. Once the addition of a new network participant is detected automatically, the coordinator starts a registration process of the new participant (cf. [3] for details).

The application within these two different setups has proven that the information contained within the metamodel is sufficient for an agent-based control of the considered aMFS. The initial effort to create the metamodel jointly with a modeling expert was quite high. We developed it over three months with a total of nine iterations (counting main metamodel versions only) and eight discussions. The metamodel's second application confirms the assumption of time-saving in the model-based development after an initial effort, as only minor adaptions (adding a parameter for providing the cost of internal aMFM routes) were necessary. Thus, the metamodel's reusability was, as to the evaluated aMFSs, considerably high.

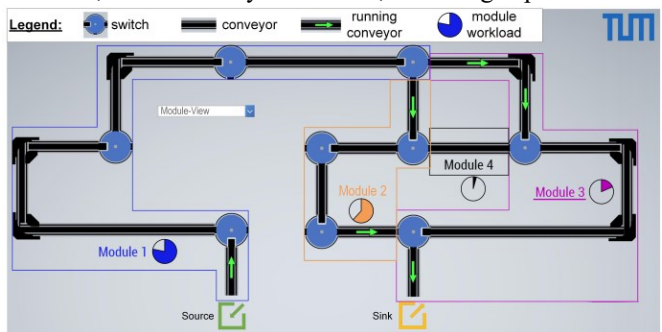

Fig. 5. Visualization of current aMFS layout highlighting contained aMFMs, their workload and the movement direction of running actuators.

As with any model, the reusability of the *IntraMAS* metamodel relates closely to its granularity: fine-grained, detailed models lead to maximum flexibility and high reusability but require high maintenance. In contrast, more generic, coarser-grained models are used when efficiency and robustness are required [34]. The developed metamodel contains abstract elements to describe properties common to all aMFMs (e.g., geometry, abilities, and interface description). These are extendable with refinements to model aMFMs with other characteristics. The abstract elements provide a uniform, module type independent, reusable structure of the required information. However, the approach is limited to stationary aMFMs handling piece goods (no modeling of bulk-material transport possible). Concerning general validity, the aim of the metamodel has to be taken into account: the representation of the information of an aMFM, which other aMFMs or the coordinator agent need, e.g., for layout calculation or route finding. In the field of logistics, aMFSs can be described as a composition of aMFMs connected via specified interfaces. The routes through these aMFSs usually consist of a sequence of connected aMFMs that contribute with at least one ability to transport a TU from source to sink. The metamodel is capable of representing these aMFMs at the logical level required to implement the agent-based control approach, and thus, taking into account its granularity, covers most of today's aMFSs.

To implement a uniform knowledge base of the module agents, a metamodel is not necessarily required if, instead, detailed programming guidelines or templates are used. However, the metamodel has proven its applicability as we have used it successfully with demonstrators and simulations.

### B. Communication Ontology for Data Exchange Between aMFM Agents

After an analysis of the information to be exchanged between the individual aMFMs within the aMFS, the control and communication are divided into two categories in accordance to [12], [22]. Some concepts propose additional communication and control categories for monitoring and offline analysis [35]. These additional categories enable extended functionality (e.g., offline optimization and remote monitoring) and will be considered in future works. The two selected categories differ concerning their real-time requirements and required update frequency. The developed communication ontology includes them both and defines the contents of the messages exchanged.

For evaluation purposes, the agent-based control approach is implemented on Beckhoff PLCs, as they support the object-oriented extensions of the IEC 61131-3 programming standard. Through this, we can encapsulate the tasks performed by the aMFM agents and their different levels. With the ADS protocol, the Beckhoff programming environment provides a simple way of establishing communication between two PLCs via TCP. The advantage of TCP communication is that the communication initiator receives feedback about the successful transmission of the data. However, technical restrictions can lead to delays in message transmission, which in the worst case even leads to the formation of a deadlock. Previous studies proved the suitability of EtherCAT to enable Plug & Produce capability even for existing network devices [36].

Additionally to information regarding pure aMFS control, data required for the HMI interface to the operator needs to be exchanged. This data is addressed in Section VI.D.

### C. Routing Process and Its Optimization

We evaluated the ssR concept using a simulation model that we implemented with the Siemens Tecnomatix Plant Simulation software. For the evaluation, a theoretical layout, as shown in Fig. 4, was used to isolate specific effects of ssRs and, additionally, we used a practical layout, which represents an aMFS in a production system, as shown in Fig. 6. Highly flexible routing concepts in aMFS are often an adaption of IP routing in communication networks or incorporate a search algorithm to find an optimal route for the current system state. Therefore, highly flexible routing concepts also choose a branch in the example in Fig. 4, dependent on the current occupancy, like routing concepts in communication networks. In contrast, the ssR concept provides knowledge about the course of future transports in the system. Due to this knowledge, a strategic optimization of ssRs, and thus of transports, can take place. The optimization aims to arrange the ssRs in order to optimize throughput and average process time. In the evaluation, we observed that the strategic optimization of ssRs assigns a dedicated branch for both, material flow relations from left to right and from right to left, and thus avoids opposing transports within one branch.

In the practical aMFS layout for a production system, we evaluated the effect of various ssRs within one system. The practical aMFS mainly consisted of conveyors, but also other aMFM types, such as a transversal carriage. At the beginning of the evaluation, only a few material flow relations were present and the shortest path from source to sink aMFM mostly determined the ssRs. With an increasing number of material flow relations, ever more areas were restricted to one-way traffic, and in some areas unidirectional loops formed. Through optimization based on ssRs, the average process time increased, but also, the throughput of the system increased, and the transports were distributed over the whole system. The evaluation showed that the decision of a TU for a path is reasonable and that ssRs provide a base for optimization, such as identifying suitable areas for a loop or establishing a one-way direction. A simulation study implemented the system layout shown in Fig. 4. When the simulation study applied ssRs with a material flow strategy, the throughput increased by about 53% compared to a control, which did not apply a strategy.

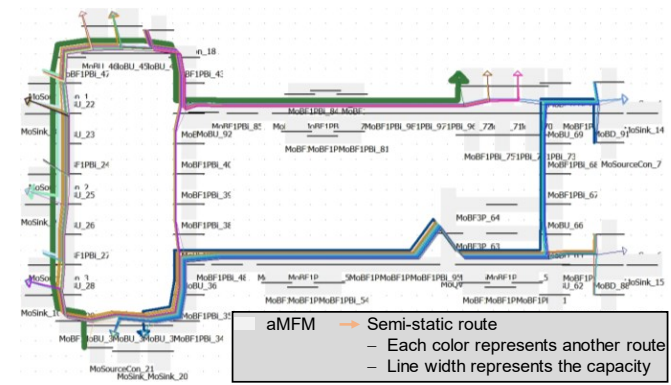

Fig. 6. Visualization of a simulation model for an aMFS in a production system with calculated and visualized semi-static routes.

### D. Visualization of aMFS Layout and MAS Decisions

We evaluated the introduced visualization concept on the MyJoghurt demonstrator. To make the MAS route decision comprehensible to the operator, we implemented a visualization with Beckhoff's TC3 HMI software, which enables platform-independent user interfaces to be developed based on current web technologies. Based on the localization data of the individual aMFMs, the visualization shows the current aMFS layout. Also, the conveying direction of the individual aMFMs, their current status, the currently active coordinator, established ssRs and alternative routes with comparison parameters (route properties) are visualized in different views. For visualization purposes, each aMFM provides its own visualization module, which consists of the 2D layout of the aMFM and its properties.

In order to generate the layout, the data required for the visualization is exchanged with each aMFM via a Beckhoff ADS interface (communication between two PLCs via TCP). However, the ADS interface has the disadvantage that aMFM agents also use the transmission channel for communication between them. To avoid the danger of disrupting the control of the aMFS, a prioritization of the information to be transmitted is required (information regarding control and visualization). Thus, we developed a concept with an OPC-UA interface as an alternative. However, OPC-UA as a standard for Industrie 4.0 does not have real-time capability with current technology.

The visualization implementation proved the general concept applicability as the aMFS layout was derived and displayed correctly. So far, we have not performed a user study regarding the visualized parameters. Also, parameters might depend on the controlled aMFS, aMFMs within and the abilities they offer.

For a summary of the considered requirements and their fulfillment through the agent-based aMFS control, cf. Table II.

### VII. Discussion and Outlook

The following section provides discussion and outlook on the performance of the MAS control concept, the derived requirements targeted within this paper and, finally, it estimates the applicability of the evaluated concept to mobile aMFMs.

### A. Effort Comparison Between Agent-based and Conventional aMFS Control

To estimate the performance of the presented MAS control approach, we compare the number of manual software changes required when adding an aMFM to an aMFS. As a benchmark, we use the highly modular, central control approach described in [2], which is very mature compared to state of the art monolithic, conventional control approaches: The PLC software is split into application-specific control logic and standardized software blocks for hardware control. A layout change requires manual adaptions of the control logic and manual insertion of layout changes, i.e., hardcoding connections of neighboring aMFMs. The average manual change effort ($CE_{Man}$) usually has to be performed under time-pressure for every layout change as described in formula (1), where $CE_{Con}$ is the change effort with conventional control after $n$ changes:

$$CE_{Con}(n) = n * CE_{Man} \quad (1)$$

In the introduced agent-based approach, the MAS performs the change of control logic automatically by registering the added aMFM's agent (implemented in standardized software bocks). The coordinator agent calculates the current layout based on the updated localization data of the aMFMs within the aMFS. Thus, layout changes do not require manual updates to the PLC software and the change effort $CE_{MAS}$ equals zero.

For the benchmark, we also consider the initial efforts: Both approaches use standard blocks for hardware control, which requires an initial development effort per module type. Thereby, the initial effort $IE_{MAS}$ of MAS standard blocks is higher than the conventional one ($IE_{Con}$), as the MAS architecture and framework have to be established in addition to pure hardware control (cf. Fig. 7). However, this resembles a onetime effort, as both are reusable for each aMFM type and aMFM-specific adaptations are negligibly small. Thus, the total effort $TE_{MAS}$ equals $IE_{MAS}$ independent of the number of layout changes $n$ according to formula (2):

$$TE_{MAS}(n) = IE_{MAS} + CE_{MAS}(n) = IE_{MAS} \qquad (2)$$

TABLE II
SUMMARY OF RESULTS

| No. | Derived Requirement | Result | Evaluation of Fulfilment | Validity |
|---|---|---|---|---|
| R1 | To enable communication and collaboration between agents, independently of their module type, the agents' knowledge base needs to be metamodel-based | True | Application with simulation and lab-sized demonstrator Successful evaluation with four reconfiguration scenarios | medium |
| R2 | The communication ontology needs to support different types of information with varying demands on real-time. | True | Separation of information to be exchanged into two types Evaluation with material flow simulation and demonstrator | high |
| R3 | A concept to enable strategic optimization based on semi-static routing is required for agent-based aMFS control. | True | Implementation of the semi-static routing concept in a simulation model (different aMFS layouts considered) Online optimization increased throughput compared to highly dynamic routing | high |
| R4.1 | Every aMFM needs to be equipped with its own visualization to enable a visualization of the current aMFS layout, including the present aMFMs and their connections. | True | Feasibility of concept demonstrated at MyJoghurt Layout visualization from bird's eye perspective is common in industry and known by operators and maintenance staff | high |
| R4.2 | The visualization of the current aMFS layout needs to be enhanced by relevant parameters to enable the operator to understand the decisions of the MAS. | Partially true | Feasibility demonstrated at lab-sized plant MyJoghurt Additional analysis required | medium |

Considering formulas (1) and (2), the total manual effort for layout changes in conventional control increases with every change, while the change effort in the presented MAS approach is equal to the initial effort $IE_{MAS}$. Therefore, after $n$ changes, the breakeven point $n_{breakeven}$ is reached, as depicted in Fig. 7.

As example of the maximum overhead in standardized blocks, we analyzed the PLC software of the hardware-wise smallest aMFM, which we used for evaluation at the MyJoghurt demonstrator. In this software block, pure hardware control makes up 66 % of the aMFM PLC software and the aMFM agent accounts for 34%. However, the high proportion of agent software results from the rather small hardware module, which requires neither much nor complex control logic. Further, this is only a rough estimation as the ratio highly depends on aspects such as the amount of comments used or the programming style.

Overall, the MAS approach requires a higher initial effort, which pays off after $n$ reconfigurations that are performable without downtime of the aMFS. Further, the MAS is developed centrally and model-based, which enhances software quality in contrast to conventional approaches, where reconfiguration with manual changes usually takes place under time pressure entailing a high risk of technical debt and low code quality. The MAS further reduces time to market in the engineering of new aMFSs by using the existing module kit, as no time-consuming development of application-specific control logic is required.

### B. Outlook Regarding Evaluated Requirements

The evaluation demonstrated the suitability of the introduced metamodel in the sense that it included all the means required to model the knowledge base of the evaluated aMFMs to be used by the aMFM agents in a uniform manner (R1). However, the metamodel's quality is not quantifiable and there might be another one, which is even more suitable for the given task. Furthermore, the evaluation was limited to a small number of different aMFM types. Thus, when including additional types, the need to adapt or enlarge the metamodel might arise. Finally, also the metamodel's application in an industrial setting will, with a high probability, lead to insights on missing elements. As we designed the metamodel in an abstract manner with the aim of being enlargeable, this does not have negative effects as long as the core structure of the metamodel remains the same. Therefore, in the next step, the application of the metamodel to model real industrial aMFSs is required to analyze whether enlargements of the metamodel are necessary to fulfill the needs of the domain entirely. Furthermore, discussions should be had with domain experts to gain insights on the feasibility of using the metamodel in real development processes. Of particular interest are approaches such as establishing an aMFM library to enhance reuse of developed aMFMs or including the model-based approach in currently deployed workflows. To further decrease the development time, data available from early phases of the development process for an aMFS should be used to instantiate the metamodel automatically, rather than implementing each aMFM agent's knowledge base manually.

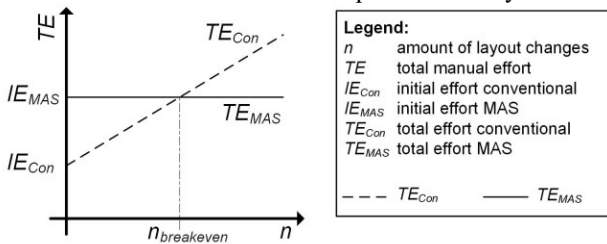

Fig. 7. Manual effort estimation in conventional and agent-based aMFS control approaches with breakeven point.

An implementation of the derived communication ontology proved its suitability to perform material transports controlled by MAS in a lab-sized demonstrator, including the real-time capable control of the material flow hardware (R2). However, the application of the concept to industrial aMFSs is required to analyze whether the derived information currently exchanged is sufficient in industrial applications or if they impose additional requirements not yet considered by the concept. Furthermore, including additional aMFM types is necessary to evaluate the extent to which the controlled processes influence the data structure of exchanged information.

Since aMFSs are similar to communication networks, MLPS for traffic optimization was adopted for semi-static routing in aMFSs. The routing concept of ssRs provides a comprehensible overview for an operator, and a basis for on- and offline traffic optimization algorithms. ssRs show a reactive behavior as they are optimized for predicted future transports. The concept of ssRs is well suited if stable material flow relations exist in the aMFS for a certain period of time. The concept of ssRs is also feasible to introduce Neural Networks (NNs) for traffic optimization in aMFS. Each set of ssRs provides a base to train a NN. A set is determined by the ssRs' properties (e.g., capacity) and the performance (e.g., throughput). In contrast, highly flexible routing determines a route for each transport dependent on all the other transports currently executed in the system. It is very complex or even impossible to isolate the effect of a route decision on the performance of the aMFS.

Regarding the visualization of the aMFS layout (R4.1), a link between module visualizations and the metamodel representing the agents' knowledge base would be beneficial as the visualized information overlaps greatly with the informational content in the metamodel. Concerning R4.2, including a visualization of the aMFM agent interaction similar to the dashboard in [37] (agent interaction in an automated warehouse) could improve understandability. Furthermore, the interaction between operator and HMI needs to be analyzed in detail, especially if the operator wants to override a decision taken by the MAS. On the one hand, it has to be discussed whether the operator is always allowed to override the MAS decisions while taking into account safety and security. On the other, manual intervention potentially requires a recalculation of routes for subsequent transportation. In a future step, the visualization concept should be applied to an industry-sized aMFS. Besides, a user study regarding the parameters and comprehensibility of the visualization should be performed to analyze the completeness of the visualized parameters.

Including concepts for the automatic updating of the visualization, such as the Module Type Package (MTP) [38], to avoid the manual addition step should be considered.

*C. Outlook Regarding Applicability to Mobile aMFMs*

The developed agent-based concept for stationary aMFMs is reusable conceptually when broadening the scope to cover mobile aMFMs, such as AGVs. Regarding the developed metamodel, general aspects of an aMFM description to be included in the agent's knowledge base can be reused: The capabilities provided by the corresponding aMFM and interfaces for material transfer need to be described in a uniform manner to enable the exchange of TUs regardless of stationary or mobile aMFMs. For this purpose, steps required and information to be exchanged in the event of a TU handover need to be modeled. The communication ontology needs to be adapted, as the information to be exchanged between mobile and stationary aMFMs differs. Nevertheless, a classification of the information in terms of their actuality requirements is also feasible for mobile aMFMs. Generally, demands on the routing process and its optimization are quite different for stationary and mobile aMFMs, as the material flow in mobile aMFMs is often identical to the route of the AGV. In this case, not the actual transport of TUs is the scope of optimization, but the route of the AGV with or without a loaded TU. Regarding visualization, mobile aMFMs impose additional requirements on the HMI due to the linkage between material flow and movement of aMFMs. However, just as in aMFSs with stationary aMFMs, establishing the trust of the operator in the MAS and enabling an operator to comprehend the MAS's decisions are of great importance. Overall, the applicability of the presented concept to mobile aMFMs is conceptually feasible in great parts but requires further analysis.

## VIII. CONCLUSION AND FUTURE WORK

The presented agent-based concept enables the development of self-configuring and flexible aMFSs. More precisely, it supports an automatic reconfiguration of the aMFS during runtime in the case of layout changes or failures, by dividing aMFSs into aMFMs, which are each controlled by a designated aMFM agent. Due to the uniform, metamodel-based knowledge base of the aMFM agents, heterogeneous aMFMs can collaborate and jointly perform transportation tasks. Additionally, the concept of ssRs, in combination with MAS, allows for an optimization of the material flow. Although agents perform the aMFS control, the routing decisions are comprehensible for the operator, as we developed a visualization concept with decisive parameters of the agent's decisions to increase the operator's trust in the MAS decisions.

For evaluation purposes, we implemented the agent-based concept on a lab-sized demonstrator, and we evaluated the scalability of the developed communication ontology successfully with a simulation study.

In future research, route optimization with Neural Networks, transport coordination in heavily loaded aMFSs and the intensive use of communication standards such as OPC-UA will be addressed. Also, it has not yet been investigated how the MAS can verify whether an aMFM agent intentionally or through a defect provides incorrect information (cf. Byzantine fault [39]). Further, the interaction between operator and MAS will be analyzed in more detail, e.g., to clarify the procedure if operators give control commands contradicting MAS decisions.

REFERENCES

[1] B. Vogel-Heuser, A. Fay, I. Schaefer, and M. Tichy, "Evolution of software in automated production systems: Challenges and research directions," *Journal of Systems and Software*, vol. 110, pp. 54–84, 2015.

[2] M. Spindler, T. Aicher, B. Vogel-Heuser, and J. Fottner, "Engineering the Control Software of Automated Material Handling Systems via Drag & Drop [Erstellung von Steuerungssoftware für automatisierte Materialflusssysteme per Drag & Drop]," *Logistics Journal: Proceedings*, 2017.

[3] C. Lieberoth-Leden, J. Fischer, J. Fottner, and B. Vogel-Heuser, "Control Architecture and Transport Coordination for Autonomous Logistics Modules in Flexible Automated Material Flow Systems," in *IEEE Int. Conf. on Automation Science and Engineering (CASE)*, Munich, Germany, 2018, pp. 736–743.

[4] T. Aicher *et al.*, "Increasing flexibility of modular automated material flow systems: A meta model architecture," *IFAC-PapersOnLine*, vol. 49, no. 12, pp. 1543–1548, 2016.

[5] D. M. Dilts, N. P. Boyd, and H. H. Whorms, "The evolution of control architectures for automated manufacturing systems," *Journal of Manufacturing Systems*, vol. 10, no. 1, pp. 79–93, 1991.

[6] P. Leitao, S. Karnouskos, L. Ribeiro, J. Lee, T. Strasser, and A. W. Colombo, "Smart Agents in Industrial Cyber–Physical Systems," *Proceedings of the IEEE*, vol. 104, no. 5, pp. 1086–1101, 2016.

[7] D. Regulin, D. Schütz, T. Aicher, and B. Vogel-Heuser, "Model based design of knowledge bases in multi agent systems for enabling automatic reconfiguration capabilities of material flow modules," in *IEEE Int. Conf. on Automation Science and Engineering (CASE)*, Fort Worth, USA, 2016, pp. 133–140.

[8] A. Albers and C. Zingel, "Challenges of Model-Based Systems Engineering: A Study towards Unified Term Understanding and the State of Usage of SysML," in *Lecture Notes in Production Engineering, Smart Product Engineering*, M. Abramovici and R. Stark, Eds., Berlin, Heidelberg: Springer Berlin Heidelberg, 2013, pp. 83–92.

[9] *System architecture for intralogistics (SAIL)*, VDI/VDMA 5100, 2016.

[10] Foundation for Intelligent Physical Agents, Ed., "FIPA Agent Management Specification," Geneva, Switzerland, 2004. Accessed: Dec. 23 2019.

[11] L. A. Cruz Salazar, D. Ryashentseva, A. Lüder, and B. Vogel-Heuser, "Cyber-physical production systems architecture based on multi-agent's design pattern-comparison of selected approaches mapping four agent patterns," *Int J Adv Manuf Technol*, vol. 8107, pp. 1–30, 2019.

[12] S. Ulewicz, D. Schütz, and B. Vogel-Heuser, "Flexible Real Time Communication between Distributed Automation Software Agents," in *Int. Conf. on Production Research (ICPR)*, 2013, pp. 1–7.

[13] J. D. Lee and K. A. See, "Trust in automation: designing for appropriate reliance," *Human Factors*, vol. 46, no. 1, pp. 50–80, 2004.

[14] S. M. Merritt, H. Heimbaugh, J. LaChapell, and D. Lee, "I trust it, but I don't know why: effects of implicit attitudes toward automation on trust in an automated system," *Human Factors*, vol. 55, no. 3, pp. 520–534, 2013.

[15] P. Marks, X. L. Hoang, M. Weyrich, and A. Fay, "A systematic approach for supporting the adaptation process of discrete manufacturing machines," *Res Eng Design*, vol. 29, no. 4, pp. 621–641, 2018.

[16] T. Beyer, P. Göhner, R. Yousefifar, and K.-H. Wehking, "Agent-based dimensioning to support the planning of Intra-Logistics systems," in *IEEE Int. Conf. on Emerging Technologies and Factory Automation (ETFA)*, Berlin, Germany, 2016, pp. 1–4.

[17] S. Libert, "Contribution for Designing Agent Based Material Flow Controls [Beitrag zur agentenbasierten Gestaltung von Materialflusssteuerungen]," Chair of Material Handling and Warehousing, TU Dortmund University, Dortmund, 2011.

[18] G. Black and V. Vyatkin, "Intelligent Component-Based Automation of Baggage Handling Systems With IEC 61499," *IEEE Trans. Automat.*

[19] A. Brusaferri, A. Ballarino, F. A. Cavadini, D. Manzocchi, and M. Mazzolini, "CPS-based hierarchical and self-similar automation architecture for the control and verification of reconfigurable manufacturing systems," in *IEEE Int. Conf. on Emerging Technology and Factory Automation (ETFA)*, Barcelona, Spain, 2014, pp. 1–8.

[20] R. Priego, N. Iriondo, U. Gangoiti, and M. Marcos, "Agent-based middleware architecture for reconfigurable manufacturing systems," *Int J Adv Manuf Technol*, vol. 92, 5-8, pp. 1579–1590, 2017.

[21] I. Kovalenko, D. Ryashentseva, B. Vogel-Heuser, D. Tilbury, and K. Barton, "Dynamic Resource Task Negotiation to Enable Product Agent Exploration in Multi-Agent Manufacturing Systems," *IEEE Robotics and Automation Letters*, vol. 4, no. 3, pp. 2854–2861, 2019.

[22] M. Vallée, H. Kaindl, M. Merdan, W. Lepuschitz, E. Arnautovic, and P. Vrba, "An automation agent architecture with a reflective world model in manufacturing systems," in *IEEE Int. Conf. on Systems, Man and Cybernetics (SMC)*, San Antonio, USA, 2009, pp. 305–310.

[23] A. Lüder, A. Calá, J. Zawisza, and R. Rosendahl, "Design pattern for agent based production system control — A survey," in *IEEE Int. Conf. on Automation Science and Engineering*, Xi'an, 2017, pp. 717–722.

[24] S. H. Mayer, "Development of a completely decentralized control system for modular continuous conveyor systems," Dissertation, Institute for Material Handling and Logistics, Karlsruhe Institute for Technology, Karlsruhe, 2009.

[25] B. Fortz and M. Thorup, "Internet Traffic Engineering by Optimizing OSPF Weights," in *Annual Joint Conf. of the IEEE Computer and Communications Societies*, Feb. 2000, pp. 519–528.

[26] D. Awduche, J. Malcolm, J. Agogbua, M. O'Dell, and J. McManus, "Requirements for Traffic Engineering Over MPLS," Network Working Group, Sep. 1999.

[27] VDI-/VDE-GMA, *Artificial Intelligence and Autonomous Systems: 10 open questions.* [Online]. Available: https://www.vdi.de/fileadmin/pages/vdi_de/redakteure/ueber_uns/fachgesellschaften/GMA/dateien/Artificial-Intelligence-and-Autonomous-Systems-10-open-Questions.pdf

[28] X. Xiao, A. Hannan, B. Bailey, and L. M. Ni, "Traffic engineering with MPLS in the Internet," *IEEE Network*, vol. 14, no. 2, pp. 28–33, 2000.

[29] M. Wilke, "Control concept for autonomous changeable material flow systems," *Logistics Journal*, 2008.

[30] J. Fischer, M. Marcos, and B. Vogel-Heuser, "Model-based development of a multi-agent system for controlling material flow systems," *Automatisierungstechnik*, vol. 66, no. 5, pp. 438–448, 2018.

[31] J. G. Carbonell, J. Siekmann, G. Goos, J. Hartmanis, J. van Leeuwen, and J. P. Müller, *The Design of Intelligent Agents*. Berlin, Heidelberg: Springer Berlin Heidelberg, 1996.

[32] A. W. t. Mors, "The world according to MARP: Multi-Agent Route Planning," Dissertation, Technische Universiteit Delft, Delft, 2010.

[33] Insitute of Automation and Information Systems, *MyJoghurt – involved in the Platform Industrie 4.0's Germany-wide online roadmap „Industrie 4.0".* [Online]. Available: http://i40d.ais.mw.tum.de/## (accessed: Dec. 23 2019).

[34] C. R. Maga, N. Jazdi, and P. Göhner, "Reusable Models in Industrial Automation: Experiences in Defining Appropriate Levels of Granularity," *IFAC Proceedings Volumes*, vol. 44, no. 1, pp. 9145–9150, 2011.

[35] Y. Luo, Y. Duan, W. Li, P. Pace, and G. Fortino, "A Novel Mobile and Hierarchical Data Transmission Architecture for Smart Factories," *IEEE Trans. Ind. Inf.*, vol. 14, no. 8, pp. 3534–3546, 2018.

[36] D. Regulin, A. Glaese, S. Feldmann, D. Schütz, and B. Vogel-Heuser, "Enabling flexible automation system hardware: Dynamic reconfiguration of a real-time capable field-bus," in *IEEE Int. Conf. on Industrial Informatics (INDIN)*, Cambridge, UK, 2015, pp. 1198–1205.

[37] D. Gürdür, K. Raizer, and J. El-Khoury, "Data Visualization Support for Complex Logistics Operations and Cyber-Physical Systems," in *Int. Joint Conf. on Computer Vision, Imaging and Computer Graphics Theory and Applications*, Funchal, Portugal, 2018, pp. 200–211.

[38] S. Wassilew, L. Urbas, J. Ladiges, A. Fay, and T. Holm, "Transformation of the NAMUR MTP to OPC UA to allow plug and produce for modular process automation," in *IEEE Int. Conf. on Emerging Technologies and Factory Automation (ETFA)*, Berlin, Germany, 2016, pp. 1–9.

[39] L. Lamport, R. Shostak, and M. Pease, "The Byzantine Generals Problem," *ACM Trans. Program. Lang. Syst.*, vol. 4, no. 3, pp. 382–401, 1982.

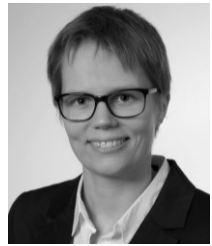

**Juliane Fischer** received an M.Sc. in Mechanical Engineering from Technical University of Munich (TUM), Munich, Germany in 2017. She is currently pursuing a Ph.D. at the Institute of Automation and Information Systems at TUM. Her main research interests are the design of modular, reusable control software and flexible material flow systems in automated Production PS.

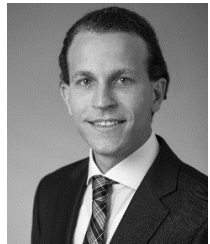

**Christian Lieberoth-Leden** received an M.Sc. in Mechanical Engineering from the Technical University of Munich (TUM), Munich, Germany in 2014. Since 2015, he works as a research assistant at the Chair of Materials Handling, Material Flow, Logistics at TUM. His main research interest is the design of modular control software for flexible automated material flow systems in automated logistics systems.

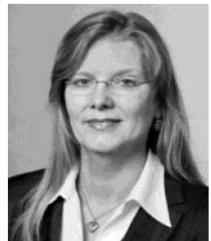

**Birgit Vogel-Heuser** received a Dr.-Ing. degree in Electrical Engineering and a Ph.D. degree in Mechanical Engineering from RWTH Aachen University, Aachen, Germany, in 1991. She was involved in industrial automation with the machine and plant manufacturing industry for nearly ten years. After holding different chairs of automation in Hagen, Wuppertal, and Kassel, she has been Head of the Automation and Information Systems Institute at the Technical University of Munich, Munich, Germany, since 2009. Prof. Vogel-Heuser was the Coordinator of the Collaborative Research Centre SFB 768: Managing Cycles in Innovation Processes—Integrated Development of Product-Service Systems Based on Technical Products. Her current research interests include systems and software engineering, and modeling distributed and reliable embedded systems.

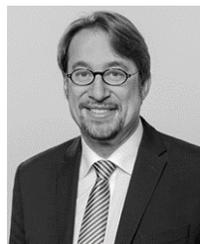

**Johannes Fottner** received a Dr.-Ing. degree in Mechanical Engineering from the Technical University of Munich (TUM), Munich, Germany in 2002. Over the past 15 years, Johannes Fottner has worked in different managing functions in the material-handling sector and gained expertise in development, implementation and operation of automated logistics and production systems composed of different modules. Since 2016, he has been the head of the Chair of Materials Handling, Material Flow, Logistics at TUM. His current research interests include investigating innovative technical solutions and system approaches to optimize logistical processes.